\documentclass[12pt]{article}
\usepackage{epsfig}
\usepackage{latexsym}
\input{amssymb.sty}

\def\bbc{{\mathbb C}}

\def\bbz{{\mathbb Z}}

\def\tr{\mathop{\rm tr}\nolimits}
\def\Tr{\mathop{\rm Tr}\nolimits}
\def\frac#1#2{{{#1}\over{#2}}}
\def\tfrac#1#2{{\textstyle{{#1}\over{#2}}}}

\def\half{\tfrac{1}{2}}
\def\third{\tfrac{1}{3}}
\def\quart{\tfrac{1}{4}}

\newcommand{\amux}{\mbox{$A_\mu (x)$}}
\newcommand{\fdown}{{\mbox{$F_{\mu\nu}$}}}
\newcommand{\fup}{{\mbox{$F^{\mu\nu}$}}}

\newcommand{\fstarup}{{\mbox{$\mbox{}^*\!F^{\mu\nu}$}}}

\newcommand{\epsup}{{\mbox{$\epsilon^{\mu\nu\rho\sigma}$}}}

\newcommand{\bq}{\begin{equation}}
\newcommand{\dq}{\end{equation}}

\begin{document}
\baselineskip 24pt
\begin{center}
{\large\bf Electric--Magnetic Duality and the Dualized Standard Model}\\
\ \\
TSOU Sheung Tsun\\
Mathematical Institute, Oxford University\\
24--29 St.\ Giles', Oxford OX1 3LB\\
United Kingdom\\
tsou\,@\,maths.ox.ac.uk
\end{center}

{\bf Abstract}\\
\baselineskip 14 pt

In these lectures I shall explain how a new-found nonabelian duality
can be used to solve some outstanding questions in particle physics.
The first lecture introduces the concept of electromagnetic duality
and goes on to present its nonabelian generalization in terms of loop
space variables.  The second lecture discusses certain puzzles that
remain with the Standard Model of particle physics, particularly aimed
at nonexperts.  The third lecture presents a solution to these
problems in the form of the Dualized Standard Model, first proposed by
Chan and the author, using nonabelian dual symmetry.  The fundamental
particles exist in three generations, and if this is a manifestation
of dual colour symmetry, which by 't Hooft's theorem is necessarily
broken, then we have a natural explanation of the generation puzzle,
together with tested and testable consequences not only in particle
physics, but also in astrophysics, nuclear and atomic physics.

Reported is mainly work done in collaboration with Chan Hong-Mo, 
and also various parts
with Peter Scharbach, Jacqueline Faridani, Jos\'e Bordes, Jakov Pfaudler, 
Ricardo Gallego severally.

\vfill
\noindent{\small Invited lectures at the 10th Oporto Meeting on Geometry,
Topology and Physics, 20--24 September, 2001, Oporto, Portugal.}

\clearpage

\section*{Introduction}
\hspace*{\parindent}In the first lecture I shall show you a new
symmetry of gauge theory which results from certain topological and
geometrical considerations.  In the second lecture I shall discuss
some theoretical problems in present-day particle physics which need
to be solved and which we think this new dual symmetry may help to 
solve.  Then in the last lecture I shall propose such a solution in
the form of the Dualized Standard Model.  What comes as a pleasant,
but intriguing, surprise is that this model has implications not only
in particle physics but also in astrophysics, nuclear and atomic
physics as well.

\subsection*{Notations, conventions and dictionary}
\hspace*{\parindent}I shall use the following groups of terms synonymously:
\begin{itemize}
\item Maxwell theory, theory of electromagnetism, abelian theory;
\item Yang--Mills theory, nonabelian (gauge) theory (however, {\em
nonabelian} may take a truly mathematical meaning);
\item Spacetime, Minkowski space.
\end{itemize}

I shall use the following notations unless otherwise stated:
\begin{itemize}
\item $X$ = Minkowski space with signature $+---$
\item $\mu,\nu,\ldots$ = spacetime indices = 0,1,2,3
\item $i,j,\ldots$ = spatial indices or group indices
\item repeated indices are summed
\item $G$ = gauge group = compact, connected Lie group\\ (usually $U(n),
SU(n), O(n)$)
\end{itemize}

I shall make the following convenient assumptions: functions,
manifolds, etc. are as well behaved as necessary; typically functions
are continuous or smooth, manifolds are $C^\infty$.

I shall use the units conventional in particles physics, in which
$\hslash =1, c=1$, the former being the reduced Planck's constant and
the latter the speed of light.

Further, since the audience is familiar with the language of fibre
bundles, it may be useful to include the following dictionary:

\smallskip

\begin{tabular}{rcl}
base space & $\longleftrightarrow$ & spacetime\\
structure group & $\longleftrightarrow$ & gauge group\\
principal bundle & $\longleftrightarrow$ & gauge theory\\
principal coordinate bundle & $\longleftrightarrow$ & gauge theory in a
particular gauge\\
connection & $\longleftrightarrow$ & gauge potential\\
curvature & $\longleftrightarrow$ & gauge field\\
holonomy & $\longleftrightarrow$&  phase factor\\
bundle reduction & $\longleftrightarrow$ & symmetry breaking\\
section $\sigma \colon X \to E$ & $\longleftrightarrow$ & Higgs
fields
\end{tabular}

\setcounter{equation}{0}

\section{Electric--magnetic duality}
\subsection{Gauge invariance}
\hspace*{\parindent}Consider an electrically charged particle in an 
electromagnetic
field.  The wavefunction of this particle is a complex-valued function
$\psi (x)$ of $X$ (spacetime).  The phase of $\psi(x)$ is not a
measurable quantity, since only $|\psi (x)|^2$ can be measured and has
the meaning of the probability of finding the particle at $x$.  Hence
one is allowed to redefine the phase of $\psi (x)$ by an arbitrary
(continuous) rotation independently at every spacetime point without
altering the physics.  This is the origin of  {\em
gauge invariance} or {\em gauge symmetry}.   Yang and Mills
generalized this phase freedom to an arbitrary element of a Lie group
(originally they considered $SO(3)$). 

In view of this arbitrariness, how can we compare the phases at
neighbouring points in spacetime?  In other words, how can we
`parallelly propagate' the phase?  Well, we know we can if given a
potential $A_\mu (x)$.  This is the connection 1-form in the principal
$G$-bundle, which is the exact geometric picture of a gauge theory.
The potential is not a directly observable quantity because it
transforms under a gauge transformation $S(x) \in G$ as:
\begin{equation}
\amux \mapsto S(x)\,A_\mu (x)\,S^{-1} (x) - (\tfrac{i}{g})\, 
\partial_\mu S(x)\,S^{-1} (x).
\end{equation}
From the connection we can define the curvature 2-form, which is the
gauge field, given in local coordinates by:
\begin{equation}
F_{\mu\nu} (x)=\partial_\nu A_\mu (x) - \partial_\mu A_\nu (x) + ig
[A_\mu (x), A_\nu (x)], \label{ymfcurl}
\end{equation}
where I have included the coupling constant $g$ whose numerical value
determines the strength of the interaction.  In abelian theory, the 6
components of this skew rank-2 tensor give exactly the 3 components of
each of the electric and magnetic fields, so that in {\em classical}
electromagnetism there is no need to introduce the potential.
However, the Bohm--Aharonov experiment demonstrates that the potential
is necessary to describe the motion of a {\em quantum} particle (e.g.\
an electron) in an electromagnetic field.  This really vindicates the
geometric description of gauge theory we have now.

In contrast to the abelian case the nonabelian field $F$ is not
observable as it is gauge covariant (that is, it is a tensorial
2-form) and not gauge invariant.  In fact, Yang has proved, and that
is what is also demonstrated in the abelian case
by the Bohm--Aharonov experiment, that it
is the set of variables comprising the holonomy of loops which
describes a gauge theory exactly.  In other words, there is a 1--1
correspondence between such sets and the sets of physical
configurations of a gauge theory.  Yang calls the holonomy the Dirac
phase factor, which is usually written in a slightly misleading
fashion as:
\begin{equation} 
\Phi (C) = P \exp ig\int_C A_\mu (x), \label{diracphase}
\end{equation}
where the letter $P$ denotes path-ordering.   We shall come back to
these loop variables later.

\subsection{Sources and monopoles}
\hspace*{\parindent}Gauge invariance in the abelian case is Maxwell's
theory of light.  To include matter, we have to look at the charges
and monopoles\footnote{This distinction is not quite accurate.  
In a fully quantized field theory, particles and fields are
synonymous.  But it is a useful and geometrically meaningful
disctinction.}.

In building a physical theory, we must look among experimental facts
to collect our ingredients.  The potential fixes only the Lie
algebra.  To select out from among the locally isomorphic ones
the correct Lie group we must look at the
particle spectrum, that is, what kind of and how many particles exist
or are postulated to exist.

Consider electrodynamics.  Since we know that all charges are
multiples (in fact, just $\pm 1$ if we do not consider composite
objects) of a fundamental charge $e$, so that
wavefunctions transform as
\begin{equation}
\psi \mapsto e^{\pm ie \Lambda} \psi,
\end{equation}
we can parametrize the circle group
$U(1)$ corresponding to the phase by $[0, 2\pi/e]$.   In fact, {\em 
charge quantization} is equivalent to having $U(1)$ as
the gauge group of electromagnetism.

On the other hand, if we consider pure electromagnetism without
charges, then the only relevant gauge transformations are those of
$A_\mu$:
\begin{equation}
A_\mu \mapsto A_\mu + \partial_\mu \Lambda, 
\end{equation}
so that the group will just be the real line given by the scalar
function $\Lambda (x)$.

Similarly for Yang--Mills theory, and for definiteness let us study an 
$\mathfrak{su} (2)$ theory.   If it contains
particles with a 2-component wave function $\psi=\{\psi_i,\ i=1,2\}$,
then
\begin{equation}
\psi \mapsto S \psi,\quad S \in SU(2),
\end{equation}
so that the effect of $S$ and $-S$ are not identical.  In this case
the gauge group is $SU(2)$.  If, on the other hand, there are no
charges so that the only gauge transformation one needs to consider 
is on the gauge potential $A_\mu (x)$:
\begin{equation}
A_\mu \mapsto S\,A_\mu\,S^{-1} - \frac{i}{g} \partial_\mu
S\,S^{-1}. 
\end{equation}
Then the effects on $A_\mu$ of $S$ and $-S$ are identical, and these
two elements should be identified, resulting in $SO(3)$ being the
gauge group.

These considerations can also be cast in terms of representations.
Charged particles in a Yang--Mills theory are in certain
representations of the gauge group.  What we are saying is the known
result that the collection of all representations determines the
group.  In the above case, the gauge potential is in the 3-dimensional
adjoint representation and the 2-component $\psi$ is in the
2-dimensional spinor representation.   In the absence of the spinor
representation, the group is $SO(3)$, but when spinors are present, the
group must be $SU(2)$.

By the same arguments, the group generally denoted $SU(3) \times SU(2) \times
U(1)$ describing the Standard Model of particle physics should really
be quotiented out by a $\bbz_6$ subgroup of its centre.   However, if
in future we either {\em discover} or {\em postulate} more
particles, then the correct group will be different.

For the moment we wish to distinguish between two types of charged 
particles: sources and monopoles.

In a pure gauge theory, that is, one without matter, we have
Yang--Mill's equation:
\begin{equation}
D_\nu F^{\mu\nu} =0, \label{ym}
\end{equation}
where $D$ denotes the covariant derivative.  {\em Electric sources}
(or just {\em sources}) are those particles that give rise to a
nonvanishing right hand side of the above equation:
\begin{equation}
 D_\nu F^{\mu\nu} = -j^\mu, \quad j^\mu= g \bar{\psi} \gamma^\mu
 \psi,
\end{equation}
where $j$ is called the current, and $\gamma^\mu$ is a Dirac gamma
matrix, identifiable as a basis element of the Clifford algebra over
spacetime. 

{\em Magnetic monopoles} (or just {\em monopoles}), on the other hand,
are topological in nature and are represented geometrically by
nontrivial $G$-bundles.  They are classified by elements of
$\pi_1(G)$.

From the definition of magnetic charges by closed curves we can easily
deduce the {\em Dirac quantization condition}.  For instance, in the
abelian case, the size of the circle is inversely proportionaly to
$e$; hence we have, in appropriate units, for $e$ electric and
$\tilde{e}$ magnetic:
\bq
e \tilde{e} =2\pi.
\dq
Similarly for nonabelian charges:
\bq
g \tilde{g} =4\pi,
\dq
the difference between the two cases being only a matter of
conventional normalization.

However, in view of the electric--magnetic duality we shall study, the
concepts of `electric' and `magnetic' are interchangeable depending on
which description one uses.  I hope to make this quite clear in the sequel.

Notice that I am using the terms `electric' and `magnetic' in a
generalized sense (that is, not just for Maxwell theory), and this
will be the case throughout this course.

\subsection{Abelian dynamics and duality: the Wu--Yang criterion}
\hspace*{\parindent}Since we are in flat spacetime, the Hodge star,
which is here more conveniently thought of as the duality operator, is
defined by
\begin{equation}
\fstarup = -\half \epsilon^{\mu\nu\rho\sigma}
F_{\rho\sigma},   \label{duality}
\end{equation}
where $\epsilon^{\mu\nu\rho\sigma}$ is the totally skew symbol 
with the convention that $\epsilon^{0123}=-1$, and the sign is a
consequence of the Minkowski signature.

It is well known that classical Maxwell theory is invariant under this
duality operator.  By this we mean that at any point in spacetime free
of electric and magnetic charges we have the two dual symmetric
Maxwell equations:
\begin{eqnarray}
\partial_\nu F^{\mu\nu}&=&0 \quad [{\rm d}\, {}^*\!F =0]
\label{freemax1}\\
\partial_\nu {}^*\!F^{\mu\nu}&=&0 \quad [{\rm d}\,F=0],
\label{freemax2}
\end{eqnarray}
where I have put in square brackets the equivalent equations in the
language of differential forms.  Then by the Poincar\'e lemma we
deduce immediately the existence of potentials $A$ and $\tilde{A}$
such that\footnote{More precisely, (\ref{fstarcurl}) follows from 
(\ref{freemax1}) and (\ref{fcurl}) from (\ref{freemax2}).}
\begin{eqnarray}
F_{\mu\nu}(x)& =& \partial_\nu A_\mu(x) - \partial_\mu A_\nu(x),
\label{fcurl}\\
{}^*\!F_{\mu\nu}(x)& =& \partial_\nu \tilde{A}_\mu(x) - \partial_\mu 
\tilde{A}_\nu(x). \label{fstarcurl}
\end{eqnarray}
The two potentials transform independently under independent gauge
transformations $\Lambda$ and $\tilde{\Lambda}$:
\begin{eqnarray}
A_\mu(x) &\longrightarrow & A_\mu(x) + \partial_\mu \Lambda(x),\\
\tilde{A}_\mu(x) & \longrightarrow & \tilde{A}_\mu(x) + \partial_\mu 
   \tilde{\Lambda}(x),
\end{eqnarray}
which means that the full symmetry of this theory is doubled to 
$U(1) \times \tilde{U}(1)$, where the tilde on the second circle group
indicates it is the symmetry of the dual potential $\tilde{A}$.  It is
important to note that the physical degrees of freedom remain the
same.  This is clear because $F$ and ${}^*\!F$ are related by an {\em
algebraic} equation (\ref{duality}).  As a consequence the physical
theory is the same: the doubled gauge symmetry is there all the time
but just not so readily detected.

This dual symmetry means that what we call `electric' or `magnetic' is
entirely a matter of choice.

In the presence of electric charges, the Maxwell equations appear
usually as
\begin{eqnarray}
\partial_\nu F^{\mu\nu}&=& -j^\mu \label{max1}\\
\partial_\nu {}^*\!F^{\mu\nu}&=&0.  \label{max2}
\end{eqnarray}
The apparent asymmetry in these equations comes from the experimental
fact that there is only one type of charges in nature which we choose
to call `electric'.  But as we see, we could equally have thought of
these as `magnetic' charges and write instead
\begin{eqnarray}
\partial_\nu F^{\mu\nu}&=& 0 \label{maxdual1}\\
\partial_\nu {}^*\!F^{\mu\nu}&=& -\tilde{\jmath}^\mu. \label{maxdual2}
\end{eqnarray}
And if both types of charges existed in nature, then we would have the
dual symmetric pair:
\begin{eqnarray}
\partial_\nu F^{\mu\nu}&=& -j^\mu \\
\partial_\nu {}^*\!F^{\mu\nu}&=& -\tilde{\jmath}^\mu.
\end{eqnarray}

This duality goes in fact much deeper, as can be seen if we use the
Wu--Yang criterion to derive the Maxwell equations\footnote{What we
present here is not the textbook derivation of Maxwell equations from
an action, but we consider this method to be much more intrinsic and 
geometric.}.

Consider first pure electromagnetism.  The free Maxwell action is:
\begin{equation}
{\mathcal A}_F^0 = - \quart \int \fdown \fup.
\label{actionff}
\end{equation}
The true variables of the theory as we said before are the $A_\mu$, so
in (\ref{actionff}) we should put in a constraint to say that $\fdown$
is the curl of $A_\mu$ (\ref{fcurl}).  This can be viewed as a
topological constraint, because it is precisely equivalent to 
(\ref{freemax2}).   Using the method of Lagrange multipliers, we form 
the constrained action
\begin{equation}
{\mathcal A} = {\mathcal A}^0_F + \int \lambda_\mu 
(\partial_\nu \fstarup), 
\label{actioncon}
\end{equation}
which we can now vary with respect to \fdown, obtaining
\bq
\fup = 2 \epsup \partial_\rho \lambda_\sigma \label{lagrange}
\dq
which implies (\ref{freemax1}).   Moreover, the Lagrange multiplier 
$\lambda$ is exactly the dual potential $\tilde{A}$.

This derivation is entirely dual symmetric, since we can equally well
use (\ref{freemax1}) as constraint for the action ${\mathcal A}_F^0$,
now considered as a functional of \fstarup:
\bq
{\mathcal A}^0_F = \quart
\int {}^*\!F_{\mu\nu} {}^*\!F^{\mu\nu},
\dq
and obtain (\ref{freemax2}) as the equation of motion.

This method applies to the interaction of charges and fields as well.
In this case we start with the free field plus free particle action:
\bq
{\mathcal A}^0 = {\mathcal A}^0_F + \int  \bar{\psi} (i 
\partial_\mu \gamma^\mu -m) \psi,
\dq
where we assume the free particle $m$ to satisfy the Dirac equation.
To fix ideas, let us consider this particle to be a magnetic
monopole.  Then the constraint we put in is (\ref{maxdual2}):
\bq
{\mathcal A}' = {\mathcal A}^0 + \int \lambda_\mu (\partial_\nu
\fstarup + \tilde{\jmath}^\mu).
\label{diracdual}
\dq
Varying with respect to $F$ gives us (\ref{maxdual1}), and varying
with respect to $\bar{\psi}$ gives
\bq
(i \partial_\mu \gamma^\mu -m) \psi = - \tilde{e} \tilde{A}_\mu
\gamma^\mu \psi.
\label{diraceom}
\dq 
So the complete set of equations for a Dirac particle in an
electromagnetic field is (\ref{maxdual1}), (\ref{maxdual2}) and
(\ref{diraceom}) if considered as a magnetic monopole.  The duals of
these equations will describe the dynamics of an electric charge, say
a positron, in an electromagnetic field.

We see from this that the Wu--Yang criterion actually gives us an
intuitively clear picture of interactions.  The assertion
that there is a monopole at a certain spacetime point $x$ means that
the gauge field on a 2-sphere surrounding $x$ has to have a certain
topological configuration (e.g.\ giving a nontrivial bundle of a
particular class), and if the monopole moves to another point, then
the gauge field will have to rearrange itself so as to maintain the
same topological configuration around the new point.  There is thus
naturally a coupling between the gauge field and the position of the 
monopole, or in
physical language a topologically induced interaction between the field
and the monopole.

As a side remark, I wish to point out that although the action
${\mathcal A}^0_F$ is not immediately identifiable as geometric in
nature, the Wu--Yang criterion, by putting the topological constraint
and the equation of motion on equal (or dual) footing, suggests that
in fact it is geometric in a subtle not yet fully understood manner.
Moreover, as pointed out, equation (\ref{lagrange}) says that the dual
potential is given by the Lagrange multiplier of the constrained
action.

\subsection{Loop space variables}
\hspace*{\parindent}We would of course like to generalize this duality
to the nonabelian Yang--Mills case.  Although there is no difficulty
in defining \fstarup, which is again given by (\ref{duality}), we
immediately come to difficulties in the relation between field and
potential, e.g. (\ref{ymfcurl}):
$$ F_{\mu\nu} (x)=\partial_\nu A_\mu (x) - \partial_\mu A_\nu (x) + ig
[A_\mu (x), A_\nu (x)].$$
First of all, despite appearances the Yang--Mills equation (\ref{ym})
$$ D_\nu F^{\mu\nu} =0 $$
and the Bianchi identity
\bq
D_\nu \fstarup =0 \label{bianchi}
\dq
are not dual-symmetric,  because the correct dual of the Yang--Mills
equation ought to be
\bq
\tilde{D}_\nu \fstarup =0,
\dq
where $\tilde{D}_\nu$ is the covariant derivative corresponding to a
dual potential.
Secondly, the Yang--Mills equation, unlike
its abelian counterpart (\ref{freemax1}), says nothing about whether
the 2-form ${}^*\!F$ is closed or not.  Nor is the relation
(\ref{ymfcurl}) about exactness at all.  In other words, Yang--Mills
equation does not guarantee the existence of a dual potential, in
contrast to the Maxwell case.   In fact, Gu and Yang have constructed
a counter-example.  Because the true variables of a gauge theory are
the potentials and not the fields, this means that Yang--Mills 
theory is {\em not
symmetric} under the Hodge star operation (\ref{duality}).

Nevertheless, electric--magnetic duality is a very useful physical
concept.  So the natural step is to seek a more general duality
transform $(\tilde{\ })$ satisfying the following properties:
\begin{enumerate}
\item $(\quad)^{\sim\sim} = \pm (\quad)$,
\item electric field $F_{\mu\nu} \stackrel{\sim}{\longleftrightarrow}
$ magnetic field $\tilde{F}_{\mu\nu}$,
\item both $A_\mu$ and $\tilde{A}_\mu$ exist as potentials (away from
charges),
\item magnetic charges are monopoles of $A_\mu$, and electric charges
are monopoles of $\tilde{A}_\mu$,
\item $\tilde{\ }$ reduces to ${}^*$ in the abelian case.
\end{enumerate}

One way to do so is to study the Wu--Yang criterion more closely.
This reveals the concept of charges as topological constraints to be
crucial even in the pure field case, as can be seen in the map below:
$$
\begin{array}{ccc}
\fbox{\shortstack{$A_\mu$ exists as \\ potential for $F_{\mu\nu}$ \\ 
$(F={\rm d}\,A)$}} & 
\stackrel{{\rm Poincar\acute{e}}}{\Longleftrightarrow}
& \fbox{\shortstack{Defining constraint \\$\partial_\mu 
\fstarup=0$ \\ $({\rm d}\,F =0)$}} \\
&&\\
\Big\Updownarrow & & \Big\Updownarrow\vcenter{%
\rlap{$\scriptstyle{\rm Gauss}$}}\\
&& \\
\fbox{\shortstack{Principal $A_\mu$ \\ bundle trivial}}  & 
 &
\fbox{\shortstack{No magnetic\\ monopole $\tilde{e}$}}\\
&&\\
{\rm GEOMETRY} && {\rm PHYSICS}
\end{array}
$$
The point to stress is that, in the above abelian case, the condition
for the absence of a topological charge (a monopole) exactly removes
the redundancy of the variables \fdown.

Now the nonabelian monopole charge was defined topologically as an
element of $\pi_1 (G)$, and this definition also holds in the abelian
case of $U(1)$, with $\pi_1(U(1)) = \bbz$.  So our first task is to
write down a condition for the absence of a nonabelian monopole.  

To fix ideas, let us consider the group $SO(3)$, whose monopole
charges are elements of $\bbz_2$, which can be denoted by a sign:
$\pm$.  The vacuum, charge $+$, that is, no monopole, is represented 
by a closed curve in the group manifold of even winding number, and
the monopole charge $-$ by a closed curve of odd winding number.  It
is more convenient\footnote{This is because sometimes it is useful to
identify the fundamental group of $SO(3)$ with the centre of $SU(2)$
and hence consider the monopole charge as an element of this centre.}, 
however, to work in $SU(2)$, which is the double
cover of $SO(3)$ and which has the topology of $S^3$.  There the
charge $+$ is represented by a closed curve, and the charge $-$ by a
curve which winds an odd number of ``half-times'' round the sphere
$S^3$.  Since these charges are defined by closed curves, it is
reasonable to try to write the constraint in terms of loop variable.
This is what we shall now study.  I must immediately say that our
analysis is not as rigorous as we would like but at the moment that is
the best we can do.  Other treatments exist but they are not so
adapted to the problem in hand.

Recall that we define the Dirac phase factor $\Phi (C)$ of a loop $C$
in (\ref{diracphase}), which can be rewritten as
\bq
\Phi[\xi] = P_s \exp ig \int_0^{2\pi} ds\,A_\mu(\xi(s))
\dot{\xi}^\mu(s), \label{phixis}
\dq
where we parametrize the loop $C$:
\bq
C: \ \ \ \{\xi^\mu(s) \colon s = 0 \rightarrow 2\pi,\ \xi(0) = \xi(2\pi)
   = \xi_0 \},
\dq
and a dot denotes differentiation with respect to the parameter $s$.
We thus regard loop variables in general as functionals of continuous
piece-wise smooth functions $\xi$ of $s$.  In this way, loop
derivatives and loop integrals are just functional derivatives and
functional integrals.   This means that loop derivatives $\delta_\mu
(s)$ are defined by a regularization procedure approximating delta
functions with finite bump functions and then taking limits in a
definite order.  For integrals, we shall ignore, for want of something
better, the question of infinite measure as usually done in physics.

\begin{figure}
\centering
\input{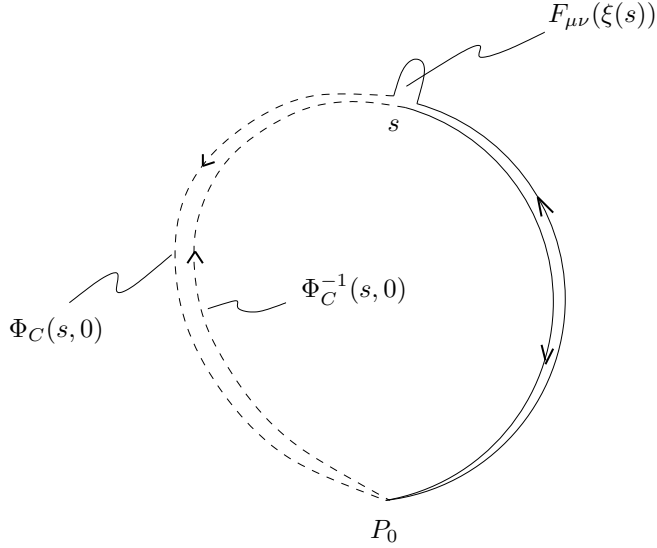}
\caption{Illustration for the quantity $F_\mu[\xi|s]$}
\label{Fmufig}
\end{figure}

Following Polyakov, we define the logarithmic loop derivative of
$\Phi[\xi]$: 
\bq
F_\mu[\xi|s] = \frac{i}{g} \Phi^{-1}[\xi]\, \delta_\mu(s) \Phi[\xi],
\label{fmu}
\dq
which acts as a kind of `connection' in loop space since it tells us
how the phase of $\Phi[\xi]$ changes from one loop to a neighbouring
loop.  We can represent it pictorially as in Figure \ref{Fmufig}.
We can go a step further and define its `curvature' in direct analogy
with $\fdown (x)$:
\bq
G_{\mu\nu}[\xi|s] = \delta_\nu(s) F_\mu[\xi|s]- \delta_\mu(s) F_\nu[\xi|s]
           + ig [F_\mu[\xi|s], F_\nu[\xi|s]].
\dq

It can be shown that using the $F_\mu[\xi|s]$ we can rewrite the
Yang--Mills action as
\bq
{\cal A}_F^0 = -\frac{1}{4 \pi \bar{N}} \int \delta \xi \int_0^{2\pi} ds
  \,{\rm Tr}\{F_\mu[\xi|s] F^\mu[\xi|s] \} |\dot{\xi}(s)|^{-2},
\label{loopaction}
\dq
where the normalization factor $\bar{N}$ is an infinite constant.
However, the true variables of our theory are still the $A_\mu$.  They
represent 4 functions of a real variable, whereas the loop connections
represent 4 functionals of the real function $\xi (s)$.  Just as in
the case of the \fdown, these $ F_\mu[\xi|s]$ have to be constrained,
but this time much more severely.  Put in another way, we have to find
the constraint on $F_\mu[\xi|s]$ in order to recover $A_\mu$ to ensure
that we are doing the same field theory as before.

It turns out that in pure Yang--Mills theory, the constraint that says
there are no monopoles:
\bq
G_{\mu\nu}[\xi|s] = 0 \label{loopcurv0}
\dq
removes also the redundancy of the loop variables, exactly as in the
abelian case.   That this condition is necessary is easy to see.  In
the absence of a toplogical charge, that is, when the principal bundle
is trivial, the potential $A_\mu$ is well-defined single-valued
everywhere.  Then the condition (\ref{loopcurv0}) follows directly
from (\ref{phixis}) and (\ref{fmu}).   The proof of the converse 
of this ``extended Poincar\'e lemma'' is fairly lengthy and will not
be presented here.   Granted this, we can now apply the Wu--Yang
criterion to the action (\ref{loopaction})
and derive the Polyakov equation:
\bq
\delta_\mu(s) F^\mu[\xi|s] = 0, \label{polyakov}
\dq
which is the loop version of the Yang--Mills equation.

In the presence of a  monopole charge $-$, the constraint
(\ref{loopcurv0}) will have a nonzero right hand side:
\bq
G_{\mu\nu}[\xi|s] = - J_{\mu\nu}[\xi|s].
\dq
The loop current $J_{\mu\nu}[\xi|s]$ has been written down explicitly,
but it is a little complicated.  However, its global form is much
easier to understand.  Recall that $F^\mu[\xi|s]$ can be thought of as
a loop connection, for which we can form {\em its} `holonomy'.  This
is defined for a closed (spatial) surface $\Sigma$ (enclosing the
monopole), parametrized by a family of closed curves $\xi_t (s),\ t=0
\to 2\pi$.  The `holonomy' $\Theta_\Sigma$ is then the total change 
in phase of
$\Phi[\xi_t]$ as $t \to 2 \pi$, and thus equals the charge $-I$.  

\begin{figure}
\centering
\input{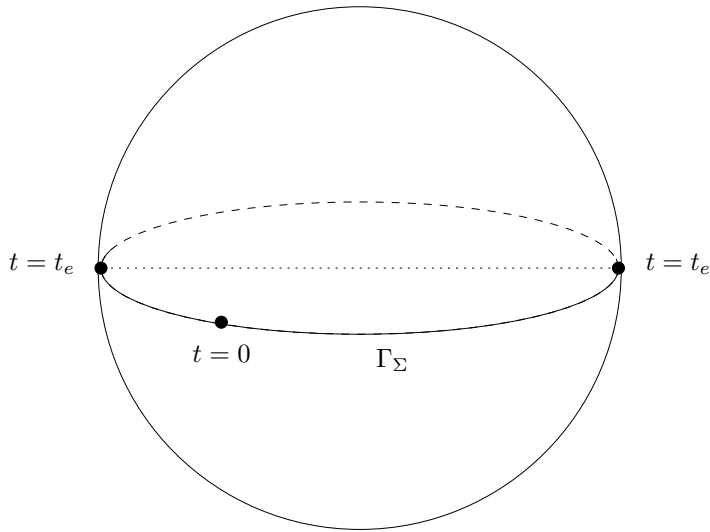}
\caption{A curve representing an $SO(3)$ monopole}
\label{holoncurv}
\end{figure}

It is instructive to examine how this result arises in detail in terms 
of the (patched) gauge potential (Figure \ref{holoncurv}).  
Without loss of generality, we shall
choose the reference point $P_0 = \xi_0^\mu$ to be in the overlap region,
say on the equator which corresponds to the loop $\xi_{t_e}$.  Starting
at $t=0$, where $\Phi^{(N)}[\xi_0]$ is the identity, the phase factor
$\Phi^{(N)}[\xi_t]$ traces out a continuous curve in $SU(2)$ 
until it
reaches $t = t_e$.  At $t = t_e$ one makes a patching transformation
and goes over to $\Phi^{(S)}[\xi_t]$.  From $t = t_e$ onwards, the phase
factor $\Phi^{(S)}[\xi_t]$ again traces out a continuous curve until $t$
reaches $2 \pi$, where it becomes again the identity and joins up with
$\Phi^{(N)}[\xi_0]$.  In order that the curve $\Gamma_\Sigma$ so traced 
out winds only half-way round $SU(2)$ while being a closed curved in 
$SO(3)$, as it should if $\Sigma$ contains a monopole, we must have
\begin{equation}
\Phi^{(N)}[\xi_{t_e}] = - \Phi^{(S)}[\xi_{t_e}],
\label{PhiNSrel}
\end{equation}
which means for the holonomy
\begin{equation}
\Theta_\Sigma = (\Phi^{(S)}[\xi_{t_e}])^{-1}\,\Phi^{(N)}[\xi_{t_e}]
=-I.
\end{equation}

\subsection{Nonabelian duality}
\begin{figure}
\centering
\input{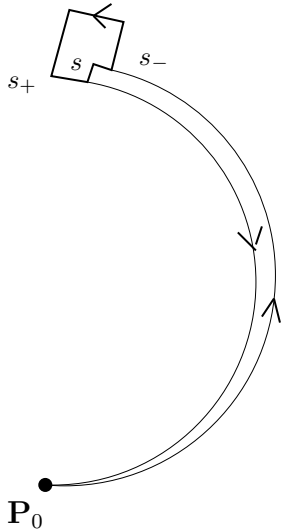}
\caption{Illustration for $E_\mu[\xi|s]$}
\label{Emuxisfig}
\end{figure}

\hspace*{\parindent}To formulate an electric--magnetic duality which
is applicable to nonabelian theory we find that we need to define yet
another set of loop variables.  Instead of the Dirac phase factor $\Phi
[\xi]$ for a complete curve (\ref{phixis}) we can define the parallel
phase transport for part of a curve from $s_1$ to $s_2$:
\bq
\Phi_\xi(s_2,s_1) = P_s \exp ig \int_{s_1}^{s_2} ds\, A_\mu(\xi(s)) 
   \dot{\xi}^\mu (s).
\dq  
Then the new variables are defined as:
\bq
E_\mu[\xi|s] = \Phi_\xi(s,0) F_\mu[\xi|s] \Phi_\xi^{-1}(s,0).
\dq
These are not gauge invariant like $F_\mu[\xi|s]$ and may not be as 
useful in general
but seem more convenient for dealing with duality.   They can be
pictorially represented as the bold curve in Figure \ref{Emuxisfig}
where the phase factors $\Phi_\xi(s,0)$ 
have cancelled parts of the faint curve representing $F_\mu[\xi|s]$.  In 
contrast to $F_\mu[\xi|s]$, therefore, $E_\mu[\xi|s]$ depends really only 
on a ``segment'' of the loop $\xi$ from $s_-$ to $s_+$.
The rule for differentiation is to take this finite segment,  
introduce the delta function variation, and
only then to take the limit as $s_- \to s_+$ in the prescribed manner.

In terms of these variables, equations (\ref{loopcurv0}) and
(\ref{polyakov}) become:
\bq
\delta_\nu(s) E_\mu[\xi|s] - \delta_\mu(s) E_\nu[\xi|s] = 0,
\label{ecurl}
\dq
and
\bq
\delta^\mu(s) E_\mu[\xi|s] = 0.  \label{ediv}
\dq

Using these new variables $E_\mu[\xi|s]$ for the field, we now define their 
`dual' $\tilde{E}_\mu[\eta|t]$ as:
\begin{eqnarray}
& & \omega^{-1}(\eta(t)) \tilde{E}_\mu[\eta|t] \omega(\eta(t)) \nonumber\\ 
&=& -\frac{2}{\bar{N}} \epsilon_{\mu\nu\rho\sigma} \dot{\eta}^\nu(t)
     \int\!\delta \xi ds\, E^\rho[\xi|s] \dot{\xi}^\sigma(s) \dot{\xi}^{-2}(s)
     \delta(\xi(s) -\eta(t)),
\label{eduality}
\end{eqnarray}
where $\omega(x)$ is a (local) rotation matrix tranforming from the frame in 
which the orientation in internal symmetry space of the fields $E_\mu[\xi|s]$ 
are measured to the frame in which the dual fields $\tilde{E}_\nu[\eta|t]$ 
are measured.  It can be shown that this dual transform satisfies all
the 5 required conditions we listed before.  Further, by
differentiating (\ref{eduality}) we can show that (\ref{ediv}) is
equivalent to 
\bq
\delta_\nu(t) \tilde{E}_\mu[\eta|t] - \delta_\mu(t)
\tilde{E}_\nu[\eta|t] = 0,  \label{edualcurl}
\dq
or that there are no monopoles in the dual field.  Hence, by the `extended 
Poincar\'e Lemma' mentioned before and via the dual of
(\ref{loopcurv0}), we deduce 
that a dual potential $\tilde{A}_\mu(x)$ will exist in this case. 

We now have a situation which is the exact parallel of pure
electromagnetism.  The equations (\ref{ecurl}) and (\ref{edualcurl})
are the equivalents of (\ref{freemax1}) and (\ref{freemax2}), and they
each guarantee as before the existence of a potential, $A_\mu (x)$ and
$\tilde{A}_\mu (x)$ respectively, again leading to a doubled gauge
symmetry $G \times \tilde{G}$.  Similarly, the Wu--Yang criterion
treats them dually: the use of one as constraint leads to the other as
equation of motion.  Here again as in the abelian case, the physical
degrees of freedom remain the same.

The treatment of charges also follows the abelian case.  To the free
action
\bq
{\mathcal A}^0= - \frac{1}{4 \pi \bar{N}} \int \Tr (E_\mu E^\mu)
\dot{\xi}^{-2} + \int \bar{\psi} (i \partial_\mu \gamma^\mu -m) \psi
\dq
we may impose the constraint
\bq
\delta_\nu(s) E_\mu - \delta_\mu(s) E_\nu = -  J_{\mu\nu},
\label{econstraint}
\dq
obtaining (\ref{ediv}) and the (dual) Dirac equation
\bq
(i \partial_\mu \gamma^\mu -m) \psi = - \tilde{g} \tilde{A}_\mu
\gamma^\mu \psi,
\dq
where the dual potential $\tilde{A}_\mu$ is again given in terms of
the Lagrange multipliers pertaining to the constraint
(\ref{econstraint}).

The whole procedure can be dualized, just as in the abelian case.  We
thus recover full electric--magnetic duality for Yang--Mills theory.

In presenting the above duality we have skipped over many technical
points and also neglected to clarify many ambiguities, because the
derivations are long and not very transparent.  It would of course be
much nicer if there is a spacetime formulation of this nonabelian
duality, but that is perhaps unlikely, particularly in view of some
recent work by Bakaert and Cucu.

\setcounter{equation}{0}

\section{Some questions in present-day theoretical particle physics}
\hspace*{\parindent}I shall not of course even try to give you a
comprehensive view of present-day particle theory.  Instead I shall
describe the background to some questions which we hope to answer
using this new-found dual symmetry\footnote{In doing so I may have
introduced some personal biases, but I shall endeavour not to do
so.}.   However, I must not give you the wrong impression that
particle theory is full of problems.  On the contrary, we now have a
very good working model of particle physics.  And it is precisely
because the theory is so good that we can now ask detailed and
profound questions.

\subsection{The Standard Model of particle physics: a first look at
the spectrum}
\hspace*{\parindent}The reason why gauge theories are so important is
that not only electromagnetism is a gauge theory, but that we now
believe that all particle interactions (that is, all except gravity)
are gauge interactions.  Moreover, all these interactions can now be
amalgamated into one single gauge theory: the Standard Model (SM).
This model knits together the two kinds of fundamental forces: strong
and electroweak\footnote{The third is gravity, which we shall not
discuss here.}.

The first thing to study is the particles themselves.  The gauge
structure provides a framework for classifying them.   Within this
framework the fundamental particles are:

\newpage
{\noindent}{\underline{Vector bosons}} (also known as gauge bosons): 
$\gamma$; $W^+,W^-,Z^0$; $g$

(photon; massive vector bosons; gluons)

\smallskip
{\noindent}{\underline{Quarks}}: $t,b$; $c,s$; $u,d$

(top, bottom; charm, strange; up, down)

\smallskip
{\noindent}{\underline{Leptons}}: $\tau,\nu_\tau$; $\mu, \nu_\mu$;
$e,\nu_e$

(tauon, tau neutrino; muon, muon neutrino; electron, electron neutrino)

\smallskip
In a full quantum theory, these particles all have corresponding
antiparticles.

The quarks and leptons are the charges we studied previously.
Experimentally, all known fundamental fermions have spin $\half$, and
all known fundamental bosons have spin 1, although theory postulates 
the existence of certain scalars called Higgs particles (of which more
later).  The quarks are not experimentally observed, but are supposed
to combine together to form most of the other particles observed in
nature and in experiments.

The quarks interact strongly and electromagnetically, while the
leptons have only electroweak interaction.   The quarks are in the
3-dimensional fundamental representation of `colour'
$SU(3)$, while the
electroweak group is usually written as $SU(2) \times U(1)$.  An
examination of all the charge assignments (as currently given) tells
us that the correct gauge group of the SM is 
$$SU(3) \times SU(2) \times U(1) / \bbz_6, $$
where $\bbz_6$ is a certain subgroup of the centre of the product
group.

The first striking feature of the above classification is that the 
charges
come in 3 copies, known as {\em generations}.  For instance, there are
the three electrically charged leptons $\tau, \mu, e$.  Except for their
very different masses: 
\bq
m_\tau \colon m_\mu \colon m_e \cong 3000
\colon 200 \colon 1 
\dq 
they have the same SM quantum numbers and 
behave in extemely similar fashions.  The 3
neutrinos $\nu_\tau,\nu_\mu,\nu_e$ also have similar interactions.
The quarks also come in 3 generations: the $U$-quarks ($t,c,u$)
with electric
charge $\tfrac{2}{3}$  and the $D$-quarks  ($b,s,d$) with electric
charge $-\third$.

The usual SM offers no explanation for the existence of 3
generations.  Moreover, experiments tell us that most probably there
are no more than 3 generations.  This is one of the major puzzles of
particle physics.

A fermion state, being an independent physical state, is orthogonal to
any other fermion states.  Hence the 3 $U$-quark states form an
orthonormal triad in generation space (which is a 3-dimensional
subspace of the total state space comprising all other quantum
numbers).  It is found experimentally that the orthonormal triad of
the $D$-quark states are not exactly aligned with that of the
$U$-quarks, and the transformation between the two triads is a unitary
matrix known as the CKM matrix.  The absolute values of the elements
of this matrix are well measured, and the off-diagonal elements vary
in magnitude from about 0.002 to 0.2.  This phenomenon is known as
`quark mixing'.   The corresponding leptonic mixing matrix, known as
the MNS matrix, is less well measured but found to have larger
off-diagonal elements.   They give rise to neutrino oscillations which
we shall study further later.

Thus a third question  that the SM does not answer is fermion mixing.

\subsection{Symmetry breaking}
\hspace*{\parindent}The strong interaction and the electroweak
interaction are both gauge theories, but they differ in one
fundamental respect.  While the strong interaction group $SU(3)$,
called `colour symmetry', is exact, the electroweak group is
`broken'.  The physical idea of symmetry breaking is that although 
the action is invariant under the action of the gauge group the
vacuum, that is, the solution to the equation of motion correspoding
to the lowest energy, is invariant under only a proper subgroup.

The gauge group of electroweak theory, when particle spectrum is taken
into account properly, is the group $U(2)$, which is doubly-covered by
$SU(2) \times U(1)$.  Both have Lie algebra $\mathfrak{su} (2) \oplus 
\mathfrak{u} (1)$.   Let $T_0$ be the generator of this $\mathfrak{u}
(1)$, and $T_1,T_2,T_3$ be those of $\mathfrak{su} (2)$, where the
notation for the $T_i$ is exactly the same as for ordinary spin, with
$T_3$ is represented by the diagonal matrix
\bq
T_3=- \frac{1}{2} \left(
\begin{array}{rr}
i & 0\\ 0 & -i 
\end{array} \right). 
\dq

The `symmetry breaking' is effected by introducing an extra term in
the Yang--Mills action:
\bq
{\mathcal A}_H = \int D_\mu \phi D^\mu \phi + V(\phi), 
\dq
with a potential $V(\phi)$ given by
\bq
V(\phi) = -\tfrac{\mu^2}{2} |\phi|^2 - \tfrac{\lambda}{4}
|\phi|^4 \quad (\lambda >0)
\dq
where $\phi$ is an $SU(2)$ doublet of  complex scalar fields, called
{\em Higgs fields}:
\bq
\phi = \left( \begin{array}{c}
\phi^+ \\ \phi^0 \end{array} \right).
\dq

Here the covariant derivative $D_\mu$ will contain four gauge components:
$W^1_\mu, W^2_\mu, W^3_\mu$ corresponding to the $\mathfrak{su} (2)$ 
part with
coupling $g_2$, and $Y_\mu$ to the $\mathfrak{u} (1)$ part with 
coupling $g_1$.

If $\mu^2 >0$, then the scalar field $\phi$ has mass $\mu$ and the
vacuum (or ground state) corresponds to $\phi_0 =0$.  If $\mu^2 <0$,
we get the famous Mexican hat potential, and the vacuum (with
$V(\phi)$ minimum) is given by
\bq
 |\phi_0| = - \mu^2/\lambda =\eta \ne 0. 
\dq
We now choose a gauge such that
\bq
\phi_0= \eta \left( \begin{array}{c}
0\\1 \end{array} \right). 
\dq
In this way, the vacuum corresponds to a particular direction in the
space of $\mathfrak{su} (2) \oplus \mathfrak{u} (1)$ and once this 
choice is
made, the physics will no longer be invariant under the whole of the
$U(2)$ group.    In fact, since 
$\phi$ is a complex vector in $\bbc^2$ , there will be a phase 
rotation left over after fixing a
direction as above, and it is  this `little group' $U(1)$
that is identified as the abelian electromagnetic group we
studied before.
Geometrically the group $U(2)$ is a torus $S^3 \times S^1$, and the
residual symmetry group is a `diagonal' $U(1)$ of this torus,
generated by the linear combination of $T_0$ and $T_3$ shown below.  

For a quantum field theory, we look at quantum excitations around the
vacuum $\phi_0$, giving rise to a new scalar field $\sigma$:
\bq
\phi(x) = \left( \begin{array}{c}
0 \\ \eta + \tfrac{\sigma (x)}{\sqrt{2}} \end{array} \right).
\dq
If we now define the Weinberg angle 
\bq
\sin \theta_W = \frac{g_1}{\sqrt{g_1^2 + g_2^2}},
\dq
and fields
\begin{eqnarray}
A_\mu & = & -\sin \theta_W\,W_\mu^3 + \cos \theta_W\,Y_\mu, 
\label{ewphoton}\\
Z_\mu & = & \cos \theta_W\,W_\mu^3 + \sin \theta_W\,Y_\mu,
\end{eqnarray}
we can re-write the action ${\mathcal A}_F^0 + {\mathcal A}_H$ in
terms of the new fields $\sigma, W^1_\mu, W^2_\mu, Z_\mu, A_\mu$.  By
comparing each term with the Klein--Gordon lagrangian for a boson with
mass $m$
\bq
-\partial_\mu \phi \partial^\mu \phi - m^2 \phi^2
\dq
we can identify the massive fields $\sigma, W^1_\mu, W^2_\mu, Z_\mu$,
while the field $A_\mu$ which remains massless we can identify as the
electromagnetic field.

To describe the charges in electroweak theory we need to introduce
further terms in the action
\bq
{\mathcal A}_L=\int \bar{\psi} D_\mu \gamma^\mu \psi + \int \rho
\bar{\psi}_L \phi \psi_R + h.c.
\label{leplag}
\dq
As a first step we shall include only one lepton generation, that is,
$e$ and $\nu_e$.  In this case,
\bq
\psi_L=\left( \begin{array}{c} \nu_e \\ e \end{array} \right)_L, 
\quad \psi_R=e_R, \quad \psi=\psi_L + \psi_R,
\dq
with
\bq
e_L= \half(1+\gamma_5)e, \quad e_R = \half (1-\gamma_5)e,
\dq
and $\nu_e$ purely left-handed\footnote{In view of the recent positive
experimental results on neutrino oscillations, this will need to be
modified.  See the next lecture.}.  In (\ref{leplag}) the second term
is called the Yukawa term (with $\rho$ a constant), and ``h.c.'' means
Hermitian conjugate.  Again by comparing (\ref{leplag}) with the Dirac
langrangian 
\bq
\bar{\psi} (i \partial_\mu \gamma^\mu -m) \psi
\dq
we can see that the electron acquires a mass through the Higgs field
$\phi$ in the Yukawa term.  The purely left-handed neutrino remains
massless in this formulation.

Thus the main result of symmetry breaking is that some gauge fields
and charges become massive.   For many theorists, the one
unsatisfactory aspect of this is that we do not have a good
theoretical reason to introduce the Higgs fields in the first place.

\subsection{The fermion mass matrices in SM}
\hspace*{\parindent}In a gauge theory under certain generally accepted
assumptions the only way for particles to be massive is by the Higgs
mechanism of symmetry breaking described above.  By confronting the
three component groups $SU(3)$, $SU(2)$, and $U(1)$ of the Standard 
Model with
what is observed (or desired to be observed), we have the following
situation as regards the mass.  Of the gauge bosons, the 8 gluons of
colour $SU(3)$ are massless, so is the particular generator of $SU(2)
\times U(1)$ corresponding to (\ref{ewphoton}) identified as the
photon.  The 3 remaining gauge bosons are massive.  Of the fermions
(which are the charges), both the quarks and the charged leptons
acquire mass through Yukawa terms involving the Higgs field.  There is
no {\em theoretical} reason to demand that the neutrinos are massless,
and indeed they most probably have a small mass.

There are 3 generations to each of the 4 types of fermions: $U,D.L,N$
(for up-type quarks, down-type quarks, charged leptons and neutrinos
respectively).
Since the left-handed components of the $U$ and $D$ quarks, as well as
those of the leptons $L$ and $N$, transform as doublets under $SU(2)$,
it is not possible in general to find a common basis in generation
space in which the 4 mass matrices are diagonal.  Indeed, because of
the observed mixing (CKM and MNS), we know that the mass matrices are
not diagonal.  Since mass is a measurable quantity, the problem is to
extract the relevant values from these matrices.

The situation is made more complicated by the fact that quantum field
theory as presently formulated can only yield measurable quantities by
a perturbative calculation, and the only realistic way to do so is by
summing Feynman diagrams.   Even putting aside the question of ghost
terms for a nonabelian gauge theory, we are immediately faced with two
problems.   Firstly each individual Feynman diagram usually contains
divergent integrals.   And even after regularizing these integrals one
has to make sure that the perturbative series can be sensibly summed.
These issues are dealt with under the heading of `renormalization' and
are definitely outside the scope of these lectures.  What we need to
know is that the renormalization procedure introduces a scale
dependence on the physical quantities in the theory.  A well-known
example is the `running coupling constant', which has a measurable effect. 

The dependence on scale $t$ of any given quantity, such as the mass
matrix, is explicitly known, via the relevant `renormalization group
equation'.   For example, for the quarks we have the first-order
SM equations:
\begin{eqnarray}
\frac{dU}{dt} &=& \frac{3}{32\pi^2} (UU^\dag - DD^\dag)U
   +(\Sigma_u - A_u)U, \label{rgeuq}\\
\frac{dD}{dt} &=& \frac{3}{32\pi^2} (DD^\dag - UU^\dag)D
   +(\Sigma_d - A_d)D,
\end{eqnarray}
where the quantities $\Sigma$ and $A$ need not concern us here.

So for a mass matrix with both eigenvalues and eigenvectors depending
on scale, it is not obvious how one can define the physical mass and
the physical state vector.   In the next lecture we shall make a
proposal for doing so in the Dualized Standard Model.

\subsection{'t Hooft's theorem and duality}
\hspace*{\parindent}There is an unexplained lopsidedness about the SM,
in that on the one hand we have an exact colour symmetry for the
strong interaction where the charges (that is, quarks) are confined
(that is, not observable in the free state), and on the other hand we
have a broken gauge symmetry for the electroweak interaction where the
charges (that is, leptons) are free.

In his famous study of the confinement problem 't Hooft introduces two
loop quantities $A(C)$ and $B(C)$ which are operators in the Hilbert
space of quantum states satisfying the commutation relation
\bq
A(C) B(C') = B(C') A(C) \exp (2\pi i n/N), \label{comrel}
\dq
for an $SU(N)$ gauge theory, where $n$ is the linking number between 
the two (spatial) loops $C$ and $C'$.  The first operator is given 
explicitly by
\bq
A(C) = \tr \Phi (C),
\dq
where $\Phi(C)$ is the Dirac phase factor (\ref{diracphase}).
He describes these two quantities as:
\begin{itemize}
\item $A(C)$ measures the magnetic flux through $C$ and creates
electric flux along $C$
\item $B(C)$ measures the electric flux through $C$ and creates
magnetic flux along $C$
\end{itemize}

So they play dual roles in the sense we have been considering in the
first lecture.  However, there was no ``magnetic'' potential available 
at the time, so
that the definition of $B(C')$ was not explicit, only through the
commutation relation above.  But we have now in fact constructed 
the magnetic potential $\tilde{A}_\mu$, and using it to construct the
dual operator $B(C)= \tr \tilde{\Phi} (C)$
we can prove the
commutation relation (\ref{comrel}), so that we know that our 
duality is the same as 't Hooft's.  This also means that we can apply
the following result to the duality we find.

\vspace*{3mm}

{\noindent}{\bf 't Hooft's Theorem.}\ \ {\em If the Wilson loop
operator of an $SU(N)$ theory and its dual theory satisfy the
commutation relation given above, then:}
\begin{eqnarray*}
SU(N)\ {\rm confined} &\Longleftrightarrow & \widetilde{SU(N)}\ 
{\rm broken}  \\
SU(N)\ {\rm broken} &\Longleftrightarrow & \widetilde{SU(N)}\ 
{\rm confined}
\end{eqnarray*}

\vspace*{3mm}

Note that the second statement follows from the first, given that the
operation of duality is its own inverse (up to sign).

The theorem does not hold for a $U(1)$ theory, where both $U(1)$ and
$\widetilde{U(1)}$ may exist in a Coulomb phase, that is, with long
range potential ($\sim 1/r$).

The statement is phrased in terms of phase transition, and has
profound implications.  It has been a cornerstone for attempts to
prove quark confinement ever since.   We shall show in the next
lecture how we use it to suggest a solution to the generation puzzle.

\smallskip

Coming back to the commutation relation, I wish to show you how to
prove it in the abelian case, just to give you a taste of what is
involved.  The nonabelian case is too complicated to treat here.

In the abelian case, we do not need the trace, hence $A(C)=\Phi(C),\
B(C')= \tilde{\Phi}(C')$, and the $\Phi$ are genuine exponentials.  So
if we can show the following relation for the exponents, we shall have
proved the required commutation relation:
$$\left[ ie \oint_C A_i dx^i,\ i \tilde{e} \oint_{C'} \tilde{A}_i dx^i
\right] = 2\pi ni. $$

Using Stokes' theorem the second integral
$$=-i\tilde{e} \int\!\!\int_{\Sigma_{C'}} {}^*\!F_{ij} d \sigma^{ij} = 
 i\tilde{e} \int\!\!\int_{\Sigma_{C'}} E_i d\sigma^i,\ {\rm where}\ 
\partial\Sigma_{C'}=C'. $$

For simplicity, suppose the linking number $n=1$.  Then the loop $C$
will intersect $\Sigma_{C'}$ at some point $x_0$---if it intersects
more than once, the other contributions will cancel in pairs, so we
shall ignore them.  So except for $x_0$, all points in $C$ are
spatially separated from points on $\Sigma_{C'}$.

Using the canonical commutation relation for $A_i$ and $E_j$
$$ [E_i (x), A_j (x')] = i \delta_{ij} \delta (x-x') $$
we get
$$\left[ ie \oint_C A_i dx^i,\ i\tilde{e} \int\!\!\int_{\Sigma_{C'}}
E_j d\sigma^j \right] = i e \tilde{e} = 2\pi i$$
by Dirac's quantization condition.

Thus we have shown explicitly in the abelian case that our definition
of duality coincides with 't~Hooft's.  The same is true in the
nonabelian case.  

\setcounter{equation}{0}

\section{The Dualized Standard Model (DSM)}
\subsection{Generation symmetry as dual colour}
\hspace*{\parindent}So far theory has made two predictions.  First, 
duality tells us that if we start with an $SU(N)$ gauge theory we 
have a doubled gauge symmetry of $SU(N) \times \widetilde{SU} (N)$.
Secondly, 't Hooft's theorem tells us that the symmetry $SU(N)$ is
confined if and only if the symmetry $\widetilde{SU} (N)$ is broken.

On the other hand, experiment gives us two pieces of information.
First, there are three and only three generations of fermions, which
are very similar except for their masses.  Secondly, $SU(3)$ colour is
confined. 

It is therefore natural, at least to us, to put the two together.  So
the main assumption of the Dualized Standard Model (DSM) is that
{\em generation symmetry is dual colour}.  In doing so, not only do we
explain the existence of exactly three generations, but also we
dispense with the need to find an experimental niche for the dual colour 
symmetry which must exist by the theory.

As a reminder, here are again the four questions we noted when
discussing the Standard Model, the first of which is now answered and
the remainder of which we shall answer:
\begin{enumerate}
\item Why are there exactly three generations?
\item What is the origin of the Higgs fields?
\item Why are the fermion masses hierarchical?
\item Why do fermions mix in the patterns observed?
\end{enumerate}

Since all fermions carry generation index, now identified as dual
colour, they are dyons in the sense of having both `electric' and
`magnetic' charges.  We have already seen the `electric' assignments
in the Standard Model.  Now a monopole of charge $n$ in SM has the
following components: $\exp (2\pi i n/3)$ of $SU(3)$, $(-1)^n$ of
$SU(2)$, $n/3$ of $U(1)$.  Putting these two together, we make the
assignments indicated in Table \ref{charges}, where bold numbers 
indicate the dimension of
the representation.  In the table, only the lightest generation is
represented, as the two higher generations have identical charges.
\begin{table}
$$\begin{array}{l|c|c|r||c|c|r|}
&SU(3) & SU(2) & U(1) & \widetilde{SU(3)} & \widetilde{SU(2)} &
\widetilde{U(1)} \\ \hline
u_L & \bf{3} & \bf{2} & 1/3 & \bf{3} & \bf{1}
& -2/3\\
d_L & \bf{3} & \bf{2} & 1/3 & \bf{3} & \bf{1}
& -2/3\\
u_R & \bf{3} & \bf{1} & 4/3 & \bf{1} & \bf{1}
& 0\\
d_R & \bf{3} & \bf{1} & -2/3 & \bf{1} & \bf{1}
& 0\\ \hline
\nu_L & \bf{1} & \bf{2} & -1 & \bf{3} & \bf{1}
& -2/3\\
e_L & \bf{1} & \bf{2} & -1 & \bf{3} & \bf{1}
& -2/3\\
\nu_R & \bf{1} & \bf{1} & 0 & \bf{1} & \bf{1}
& 0\\
e_R & \bf{1} & \bf{2} & -2 & \bf{1} & \bf{1}
& 0\\ \hline
\end{array}$$
\caption{Charge assignments in DSM}
\label{charges}
\end{table}

\subsection{Higgs fields as frame vectors}
\hspace*{\parindent}Next we recall that, in the duality transform
(\ref{eduality}), $\omega(x)$  was originally 
conceived as the matrix
relating the internal symmetry $G$-frame to the dual symmetry
$\tilde{G}$-frame.  The rows of $\omega$ therefore transform as the 
conjugate fundamental representation of the $G$-symmetry, i.e. as
$\bar{\bf 3}$ of colour or $\bar{\bf 2}$ of weak isospin, while its 
columns transform as the fundamental representation of the dual 
$\tilde{G}$-symmetry, i.e. as ${\bf 3}$ of dual colour or ${\bf 2}$ of 
dual weak isospin.  We want to relate these to the Higgs fields.

The 
idea of using frame vectors as dynamical variables is made familiar 
already in the theory of relativity where in the Palatini treatment or 
the Einstein-Cartan-Kibble-Sciama formalism the space-time 
frame vectors or vierbeins are used as dynamical variables.  In gauge 
theory, frame vectors in internal symmetry space are not normally given 
a dynamical role, but it turns out that in the dualized framework they 
seem to acquire some dynamical properties, in being patched, for example, 
in the presence of monopoles.  Moreover, they are 
space-time scalars belonging to the fundamental representations of the 
internal symmetry group, i.e. doublets in electroweak $SU(2)$ and 
triplets in dual colour $\widetilde{SU}(3)$, and have finite lengths 
(as vev's).  They thus seem to have just the right properties to be 
Higgs fields, at least as borne out by the familiar example of the 
Salam-Weinberg breaking of the electroweak theory.  Hence, one makes 
the second basic assumption in the DSM scheme, namely that these frame 
vectors are indeed the physical Higgs fields required for the spontaneously 
broken symmetries.

Having made these basic assumptions, let us now explore the consequences.
First, making the frame vectors in internal symmetry space into dynamical
variables and identifying them with Higgs fields mean that for dual
colour $\widetilde{SU}(3)$, we introduce 3 triplets of Higgs fields
$\phi^{(a)}_a$, where $(a) = 1, 2, 3$ labels the 3 triplets and $a =
1, 2, 3$ their 3 dual colour components.  Further, the 3 triplets having
equal status, it seems reasonable to require that the action be symmetric 
under their permutations, although the vacuum need not be.  
An example of a Higgs potential which breaks both this permutation symmetry 
and also the $\widetilde{SU}(3)$ gauge symmetry completely is as follows: 
\begin{equation}
V[\phi] = -\mu \sum_{(a)} |\phi^{(a)}|^2 + \lambda \left\{ \sum_{(a)}
   |\phi^{(a)}|^2 \right\}^2 + \kappa \sum_{(a) \neq (b)} |\bar{\phi}^{(a)}
   .\phi^{(b)}|^2,
\label{Vofphi}
\end{equation}
a vacuum of which can be expressed without loss of generality in terms of 
the Higgs vacuum expectation values:
\begin{eqnarray}
\phi^{(1)} = \zeta \left(
\begin{array}{c}
 x \\ 0 \\ 0 
\end{array} \right), \,\,\,
\phi^{(2)} = \zeta \left(
\begin{array}{c}
 0 \\ y \\ 0 
\end{array} \right) , \,\,\,
\phi^{(3)} =  \zeta \left(
\begin{array}{c} 
0 \\ 0 \\ z 
\end{array} \right),
\label{vevs}
\end{eqnarray}
with
\begin{equation}
x^2 + y^2 + z^2 = 1,
\label{normvevs}
\end{equation}
and
\begin{equation}
\zeta = \sqrt{\mu/2 \lambda},
\label{zeta}
\end{equation}
$x, y, z$, and $\zeta$ being all real and positive.  Indeed, this vacuum
breaks not just the symmetry $\widetilde{SU}(3)$ but the larger symmetry 
$\widetilde{SU}(3) \times \tilde{U}(1)$ completely giving rise to 9 massive 
dual gauge bosons.  And of the 18 real components in $\phi^{(a)}_a$, 9 are 
thus `eaten up', leaving just 9 (dual colour) Higgs bosons.

\subsection{The fermion mass matrices}
\hspace*{\parindent}Following the procedure outlined in the last 
lecture, we let the
fermions acquire nonzero mass through additional terms in the
Lagrangian.  It turns out that, using frame vectors as Higgs fields as
above, 
the $3 \times 3$ fermion mass matrix
(one for each of the four types $T = U, D, L, N$) is of rank 1 and can
thus be written in terms of one single normalized vector $v_T=(x,y,z)$
(without loss of generality we may assume $x \geq y \geq z$) as:
\bq
m=m_T \left( \begin{array}{c}
x\\y\\z \end{array} \right) 
(x\ y\ z) = m_T \left( \begin{array}{ccc}
x^2 & xy & xz\\
xy & y^2 & yz \\
xz & yz & z^2
\end{array} \right),
\label{massmat}
\dq
where $m_T$ is the only nonzero eigenvalue.

\begin{figure}
\centerline{\psfig{figure=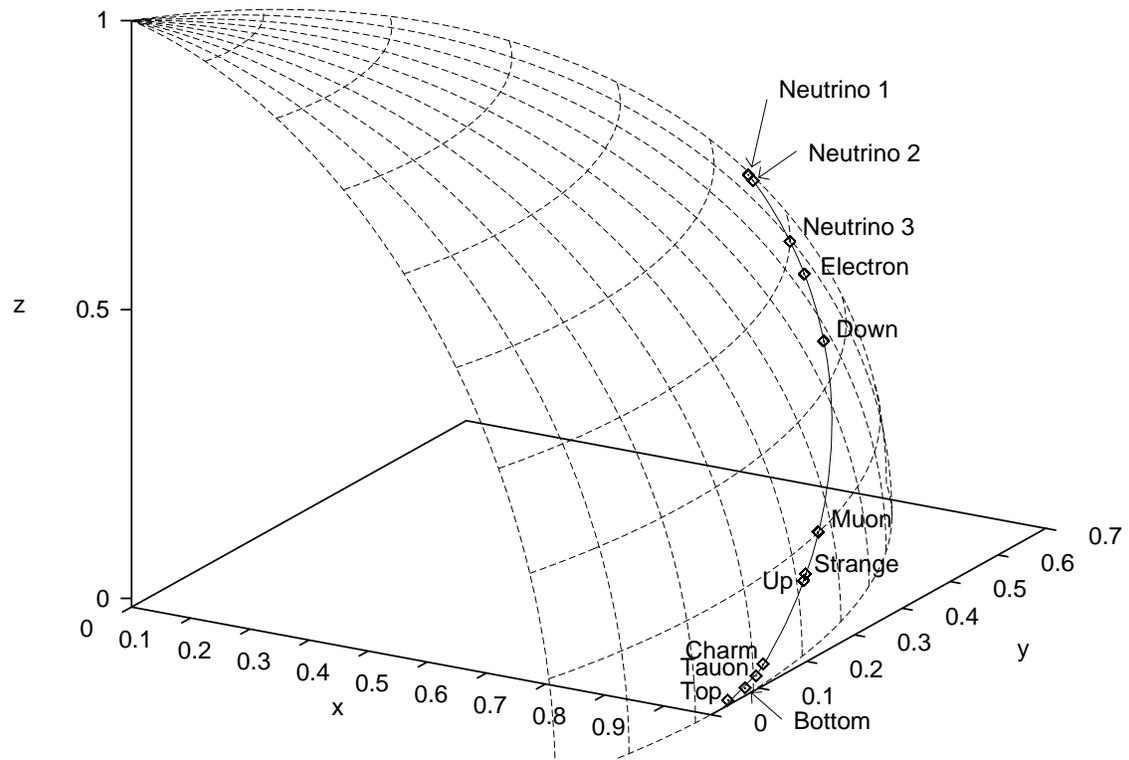,width=0.85\textwidth}}
\caption{The trajectory traced out by $v_T$ as scale changes}
\label{sphere}
\end{figure}

Evaluating the 1-loop Feynman diagrams in the standard way gives us the
renormalization group equation for $m$.  We can show that the matrix
$m$ remains of factorized form under change of scale $\mu$, so that
all the physics content of $m$ can be deduced from the running of the
normalization factor $m_T$ and the rotation of the unit vector $v_T$.
In particular, we have the following renormalization group equation for
$v_T$: 
\bq
\frac{d}{d(\ln \mu^2)} \left( \begin{array}{c} x \\ y \\ z 
   \end{array} \right)
   =  \frac{3}{64 \pi^2} \rho^2 \left( \begin{array}{c}
         \tilde{x} \\ \tilde{y} \\ \tilde{z} \end{array} \right),
\label{ourrge}
\dq
with 
\bq
\tilde{x} = \frac{x(x^2-y^2)}{x^2+y^2} + \frac{x(x^2-z^2)}{x^2+z^2},
   \ \ \ {\rm cyclic},
\dq
where $\rho$ is constant, the same for all types $T$, which gives the
`speed' at which the vector $v_T$ runs.  This means that all 
$v_U, v_D, v_L, v_N$ lie on the same renormalization group equation
trajectory on the unit sphere, Figure \ref{sphere}, a study of which
will give us most of the results we want.

Recall from ({\ref{massmat}) that the mass matrix $m$ has only one
nonzero eigenvalue, but both it and the eigenvectors depend on the
scale $\mu$.  The eigenvector corresponding to the single nonzero
eigenvalue is $v_T$, the radial vector of the trajectory on the unit
sphere (Figure \ref{sphere}).  Further, we can easily see that the
renormalization group equation (\ref{ourrge}) has two fixed points:
$v=(1,0,0)$ at high (or infinite) energy, and
$v=(1/\sqrt{3},1/\sqrt{3},1/\sqrt{3})$ at low (or zero) energy.
These are important for what follows.

We raised the question in the last lecture about the ambiguity in
defining physical masses and states when the mass matrix runs with
scale.  We now make the following proposal, a kind of working
criterion which is applicable in DSM.   

First we run $m$ to a scale $\mu$ such that $\mu=m_T(\mu)$; this value
we can reasonably call the mass of the highest generation.  To fix
ideas, let us concentrate on the $U$ type quarks $t,c,u$.  The
corresponding eigenvector $v_t$ is then the state vector of the
$t$ quark.  Having fixed this, we now know that the $c$ and $u$ lie in
the 2-dimensional subspace $V$ orthogonal to $v_t$.  As we go down in
scale the eigenvector $v_T$ rotates so that $V$ is no longer the null
eigenspace, and the projection of $m$ onto $V$ is a $2 \times 2$
matrix, again of rank 1.  We now repeat the procedure for this
submatrix and determine both the mass and the state of the $c$ quark
(Figure \ref{quarkstates}).
\begin{figure}
\centerline{\psfig{figure=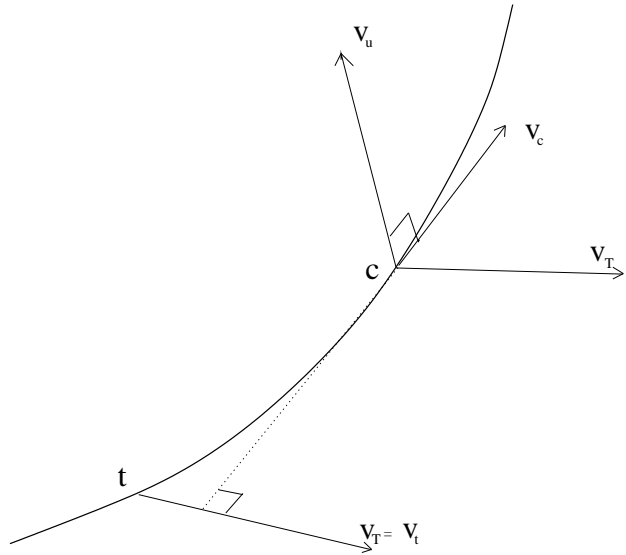,width=0.6\textwidth}}
\caption{The state vectors of the 3 physical states of the $U$ type quark.}
\label{quarkstates}
\end{figure}
Once the $c$ state is determined, we know the $u$ state as well, as
being the third vector of the orthonormal triad $(t,c,u)$.  The $u$
mass is similarly determined.

The above procedure can be repeated for the other three types of
fermions\footnote{Neutrinos will need further special treatment.  See
next subsection.}.  The direction cosines of the two triads $(t,c,u)$ and
$(b,s,d)$, in other words, the 9 inner products, will give us the CKM
matrix, Figure \ref{quarktriads}.
\begin{figure}
\centerline{\psfig{figure=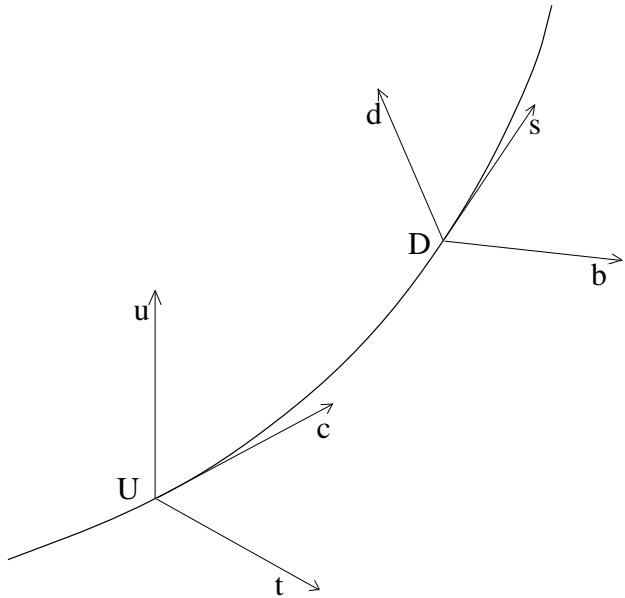,width=0.6\textwidth}}
\caption{Two triads of state vectors for the quarks on the trajectory}
\label{quarktriads}
\end{figure}
Similarly for the MNS mixing matrix of the leptons.

Our proposal has the following desirable properties:
\begin{itemize}
\item The 3 lepton state vectors (of the same type) are always orthogonal.
\item The mixing matrix is always unitary.
\item Mass hierarchy is automatic.
\end{itemize}
The last point comes about because the lower generation masses are
obtained by the ``leakage'' from the single nonzero eigenvalue of $m$
after the heaviest generation is fixed.

This now answers the remaining two of our four questions.

\subsection{Neutrino masses}
\hspace*{\parindent}Before we present the consequences of DSM, we need
to make a small digression on neutrinos.  Recall that in SM the
neutrinos have only left-handed components, so that one cannot write a
Yukawa term $\bar{\nu}_L \phi \nu_R$ as in (\ref{leplag}).  However,
as we emphasized, there is no theoretical reason why right-handed
neutrinos should not exist.

In fact, in view of the recent neutrino oscillation results from
Super\-kam\-io\-kande, SNO and others, evidence is quite conclusive that
neutrinos must have nonzero masses, although these are very small
compared to the other fermions.

Neutrinos oscillate because the states $\nu_\tau,\ \nu_\mu,\ \nu_e$ in
which they are produced, for example in $\beta$-decay where a neutron
decays into a proton, an electron and a neutrino, are not mass
eigenstates (that is, states which propagate according to the Dirac
equation).

The most naive way of just introducing a right-handed component
$\nu_R$ is not very satisfactory, because of their very small masses.
One way to produce very small physical neutrino masses is via the
`see-saw mechanism'.   Because $\nu_R$ has no gauge charges whatsoever
(in SM) there is an additional mass term, the Majorana mass term, one
can write down.  By postulating a large Majorana mass $B$ (which has
no counterpart in the other fermions), the see-saw mechanism can
produce a small $\nu$ mass when the following matrix is diagonalized:
$$\left(
\begin{array}{cc}
0 & M \\ M & B \end{array} \right),
$$
where $M$ the Dirac mass (coming from the Yukawa term) need not be too
different from that of the other fermions, and yet one can have a
small eigenvalue $\sim M^2/B$.

Now DSM can quite naturally incorporate this feature and this is what
has been done.

\subsection{Consequences of DSM: masses and mixing}
\hspace*{\parindent}As the audience is mainly mathematical, I 
shall mention only very
briefly the numerical results.  We used the so-called Cabibbo angles
$V_{us}$ and $V_{cd}$, which are the best measured experimentally
among the mixing angles, as inputs.  So for the quark CKM matrix:
$$
\left( \begin{array}{lll}
V_{ud} & V_{us} & V_{ub}\\
V_{cd} & V_{cs} & V_{cb}\\
V_{td} & V_{ts} & V_{tb} \end{array} \right)
$$
with experimental values
\bq
\left( \begin{array}{lll} 
       0.9742 - 0.9757 & 0.219 - 0.226 & 0.002 - 0.005 \\
       0.219 - 0.225 & 0.9734 - 0.9749 & 0.037 - 0.043 \\
       0.004 - 0.014 & 0.035 - 0.043 & 0.9990 - 0.9993 \end{array} \right)
\label{ckmexp}
\dq
we obtain 
\bq
\left( \begin{array}{ccc} 
              0.9752 & 0.2215 & 0.0048 \\
              0.2211 & 0.9744 & 0.0401 \\
              0.0136 & 0.0381 & 0.9992 
              \end{array} \right)
\label{ckmthe}
\dq
from our calculations.  And for the lepton MNS matrix:
$$
\left( \begin{array}{lll}
U_{e1} & U_{e2} & U_{e3}\\
U_{\mu 1} & U_{\mu 2} & U_{\mu 3}\\
U_{\tau 1} & U_{\tau 2} & U_{\tau 3} \end{array} \right)
$$
with experimental values
\bq
\left( \begin{array}{ccc}
       \ast & 0.4 - 0.7 & 0.0 - 0.15 \\
       \ast & \ast & 0.56 - 0.83 \\
       \ast & \ast & \ast \end{array} \right)
\label{mnsexp}
\dq
we obtain
\bq
\left( \begin{array}{ccc} 
              0.97 & 0.24 & 0.07 \\
              0.22 & 0.71 & 0.66 \\
              0.11 & 0.66 & 0.74 
              \end{array} \right)
\label{mnsthe}
\dq
from our calculations.

All except the so-called `solar neutrino angle' $U_{e2}$ are well
within experimental bounds.  We shall now give a simple geometric
reason for such good agreements, quite apart from the actual results
of our numerical computation.

Since we have a curve on the unit sphere, we are reminded of the
elementary classical differential geometry of a curve on a surface.
Let $N$ be the normal to the surface at a given point,  $T$ the 
tangent to the curve at that point, and $B=N \wedge T$.  These 3 vectors
form the Darboux triad, and the rotation matrix linking two
neighbouring triads at $\Delta s$ apart are given by the 
Serret--Frenet--Darboux formulae\footnote{These are similar to the
well-known Serret--Frenet formulae for a space curve, with tangent,
normal and binormal forming a moving triad.} as
\bq
\left( \begin{array}{ccc}
       1 & -\kappa_g \Delta s & -\tau_g \Delta s \\
       \kappa_g \Delta s & 1 & \kappa_n \Delta s \\
       \tau_g \Delta s & -\kappa_n \Delta s & 1 \end{array} \right),
\label{sfdgen}
\dq
where $\kappa_g$ is the geodesic curvature, $\kappa_n$ the normal 
curvature, and $\tau_g$ the geodesic torsion.  Now for a unit sphere, 
$\kappa_n=1,\ \tau_g=0$, so that we have for the  case in hand
\bq
\left( \begin{array}{ccc}
       1 & -\kappa_g \Delta s & 0 \\
       \kappa_g \Delta s & 1 & \Delta s \\
       0 & -\Delta s & 1 \end{array} \right).  
\label{sfd}
\dq

A comparison of (\ref{sfd}) with (\ref{ckmexp}) (or indeed
(\ref{ckmthe})) and with (\ref{mnsexp}) (or again (\ref{mnsthe})) will
elicit the following remarkable features immediately:
\begin{enumerate}
\item The corner elements ($V_{ub},\ V_{td}$) and respectively
($U_{e3},\ U_{\tau 1}$) are at most second order in the separation
$\Delta s$, and hence vanishingly small.
\item The 4 other off-diagonal CKM elements are small compared with
the diagonal elements, since they are of first order in the separation
between the $t$ and the $b$ quarks, which is small as seen in 
Figure~\ref{sphere}. 
\item The elements $V_{cb}, V_{ts}$ for quarks are much smaller
than their counterparts $U_{\mu 3}, U_{\tau 2}$ for leptons, since they
are to first order proportional to the separation, which is much smaller
for quarks than for leptons as seen in Figure~\ref{sphere}.
\end{enumerate}
Incidentally, we also understand why it is more difficult to get the
solar angle $U_{e2}$ correct, since this depends more on the details
of the actual curve, namely, its geodesic curvature.  (The
corresponding quark elements were inputs, as already mentioned.)

What is quite amusing is that, although the $\tau$ and $\nu_3$ are
quite far apart on our trajectory (Figure~\ref{sphere}), the relation
(\ref{sfd}) still seems to work, and in fact if you take a piece of
string and measure the ratio of the distance between $\tau$ and
$\nu_3$ and between $t$ and $b$, you will get roughly the correct
factor $U_{\mu 3}/V_{cb} \sim 20$.  This must be the most inexpensive
particle physics experiment ever performed!

\subsection{Renormalization fixed points}
\hspace*{\parindent}In determining the mixing angles the fermion
masses necessarily come into play, as explained in \S 3.3.  However, I
did not explicitly present the mass results in the last subsection,
because these are much more transparent when examined from the point
of view of the renormalization group equation fixed points.

Recall that we need only concern ourselves with the single unit vector
$v_T$.  As the scale $\mu$ decreases, this vector rotates from near
the high energy fixed point at $(1,0,0)$ to the low energy fixed point
 $(1/\sqrt{3},1/\sqrt{3},1/\sqrt{3})$ tracing out a trajectory
(Figure~\ref{sphere}).   Figure~\ref{thetaplot}
\begin{figure}
\centering
\hspace*{-2.2cm}
\input{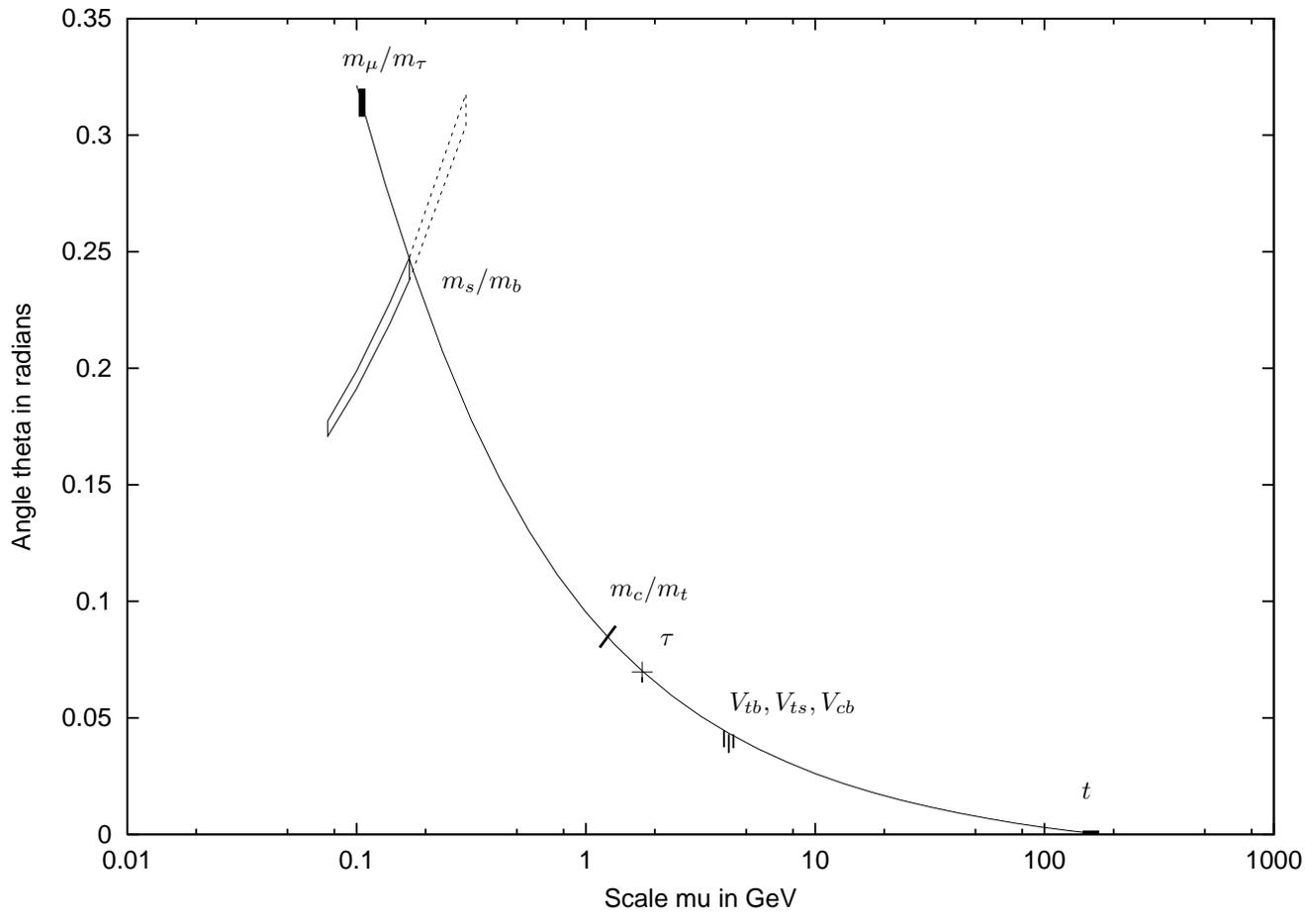}
\caption{Trajectory near the high energy fixed point}
\label{thetaplot}
\end{figure}
shows that part of the trajectory which is near the high energy fixed
point, and which is very nearly planar.  Since we are near a fixed
point, rotation is slow in the sense that (i) the top quark at 175 GeV
is almost indistinguishable from the fixed point itself, and (ii)
after running down several decades of energy the rotation angle is
still fairly small.  That the masses and mixing parameters come from a
rotation angle which remains small even when the energy range is large
is very important for the DSM results.  This is because our
renormalization group equation comes from contributions from 1-loop
diagrams only and if the relevant parameter (here the rotation angle)
were not small there would be no reason to expect 1-loop calculations
to be accurate over such a large range in scale.  We see in
Figure~\ref{thetaplot} that the DSM curve here goes smoothly
through all the data points (mass and mixing).   Notice that 
experimental estimates of quark masses
are not without ambiguity, hence the large `error bars' on the data
points.  

Since all three state vectors (in generation space) of the three
generations are already fixed at the position of the second
generation, Figure~\ref{thetaplot} tells us that they are accurately
determined (for both quark types $U$ and $D$ and for charged leptons
$L$), hence good agreement with experiment for the CKM mixing
elements.  To get the lowest generation masses $u,d,e$ one needs to run
further down in energy, where the rotation angles become sizeable.
While the values we get are still sensibly hierarchical (a necessary
consequence of our scheme, \S 3.3) they are numerically
inaccurate\footnote{These were obtained about 3 years ago.  We now
believe that we may have a more reliable method of extrapolation
which would give closer agreement with experiment, especially for the
$u$ and $d$ quarks.}.

At the other extreme, the neutrinos have very small mass.  The
heaviest one $\nu_3$ is estimated to be $\sim 0.05$ eV, which is so
near the low energy fixed point that its state vector is very well
approximated by the fixed point itself
$(1/\sqrt{3},1/\sqrt{3},1/\sqrt{3})$, as confirmed by explicit
calculations. 
Since we already know the
charged lepton triad accurately, we can now deduce the `atmospheric'
angle $U_{\mu 3}$ and the `Chooz' angle $U_{e3}$, the former near
maximal and the latter very small, again as confirmed by calculations
and exactly as data say.  The `solar' angle $U_{e2}$, however, depends
on how the trajectory approaches the fixed point, that is, its tangent
there, which we cannot obtain simply by extrapolation.
For this angle, we do not get good agreement with experiment (see also
the last subsection).

\subsection{Further consequences of DSM}
\hspace*{\parindent}To summarize so far: with 3
parameters (fixed from data) we were able to calculate all the other
quantities relevant to fermion masses and mixing.  Of these, 9 are
within experimental error, 1 nearly so, and only 2, the
electron mass and the solar neutrino angle are outside experimental
bounds. 

With all but one piece (see below) of information in hand we can now
look at DSM in three further broad areas of application.

First, the exchange of dual colour gauge bosons will induce new
interactions, called `flavour-changing neutral current effects'.  By
assuming a lower bound of about 500 TeV for the mass of these dual
colour bosons, we are able to satisfy experimental bounds for the
following rather diverse interactions: 
\begin{itemize}
\item rare hadron decays e.g. $K_L \to e^\pm \mu^\mp$;
\item mass differences e.g. $K_L - K_S$;
\item coherent muon-electron conversion on nuclei e.g. $ \mu^- + Ti
\to e^- + Ti$;
\item muonium conversion e.g. $\mu^+ e^- \to e^+ \mu ^-$;
\item neutrinoless double beta decay e.g. ${}^{76}Ge \to {}^{76}Se + 2 e^-$.
\end{itemize}
Note that these involve not only particle physics but also nuclear and
atomic physics.

Secondly, a rotating lepton mass matrix will mean that lepton quantum
numbers may not be conserved, leading to phenomena of lepton flavour
violation or `transmutation'.  Using the parameters determined 
previously we have
done calculations for the following interactions:
\begin{itemize}
\item decays e.g. $\Upsilon \to \mu^\pm \tau^\mp$;
\item photo-transmutation e.g. $\gamma e^- \to \gamma \tau^-$;
\item transmutational Bhabha e.g. $e^+ e^- \to e^+ \mu^-$;
\end{itemize}
all predicted cross-sections satisfying existing bounds.

Lastly, and rather surprisingly, DSM has something to say about the
so-called `ultra high energy cosmic rays' or `airshowers'.  These are
still an unsolved puzzle in astrophysics, and I shall describe them
briefly, as they are something of an expected bonus for us.

Cosmic rays with energy $> 10^{20}$ eV pose a problem in
astrophysics.  Over the last 30 years about 12 such events have been
observed, each producing some $10^{11}$ charged particles.  If they
are protons, they will lose their energy quickly by interacting with
the cosmic microwave backgrouns: $ p +\gamma_{2.7} \to
\Delta + \pi$.  Greisen, Zatsepin and Kuz'min (GZK) estimated that if
these primary particles are indeed protons, they cannot therefore
originate further than 50 Mpc away without losing their energy.
However, there are no obvious proton sources of such high energy
that near.   Moreover, some possible pairs and triples have been
observed, pointing back to the same source.  But if they were protons,
they would have been deflected by the inter-galactic magnetic field
and one would not have been able to trace back their origin.  So
protons seem not to provide a solution.
If these
particles are neutrinos, on the other hand, they would not suffer from
these constraints.  
But ordinary weakly interacting neutrinos would not have a
large enough cross-section with air nuclei to produce the many
particles observed in each such event.

DSM offers a possible solution.  At energies above the dual colour
gluon mass neutrinos will have become strongly interacting, because
they carry the generation index which is identified with dual colour.
From the GZK bound we can deduce a lower bound for the mass of these
bosons, which turns out to be around 500 TeV.   That this lower bound
coincides with the upper bound estimated from flavour-changing neutral
currents above (more specifically, from $K_L - K_S$ mass difference)
is a very pleasant surprise!   Indeed, strongly interacting
neutrinos in this scenario can actually solve all the above-mentioned
problems: 
\begin{itemize}
\item $\nu$ can escape strong electromagentic field around any 
source, e.g. AGN such as MCG8-11-11;
\item $\nu$ can survive a long journey through microwave background;
\item near hadronic cross-section with air nuclei at high energy;
\item pairs (or triplets) not deflected by inter-galactic
electromagnetic  field;
\item highest energy event at $3 \times 10^{20}$ eV with no abundant
lower energy events in same direction: $\nu$ interacts strongly only at high
energy.
\end{itemize}
Indeed, for this last event from Fly's Eye we estimated the height of
the primary vertex and found it agrees substantially better with a
neutrino-induced rather than a proton-induced shower, 
Figure~\ref{maxangle}.
\begin{figure}
\centerline{\psfig{figure=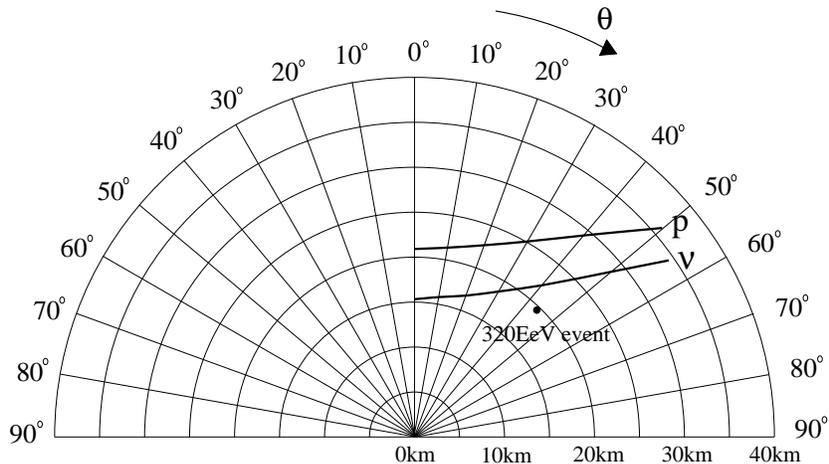,width=0.8
\textwidth}}
\vspace{1cm}
\caption{The positions of the maximum heights for varying $\theta$}
\label{maxangle}
\end{figure}

More similar quantitative calculations within DSM can be done.
With the planned Auger observatories, we hope that the question about
the origin of airshowers will be settled by the new data in perhaps
the next decade.

\subsection{DSM: the future}
\hspace*{\parindent}We have travelled a long way.  From the loop space
formulation of gauge theory, by way of electric--magnetic duality and
't~Hooft's theorem, we have arrived at actual numbers to confront
with experiments and have so far done honourably indeed.  But many
things remain to be understood and done.

While the basis of DSM seems to have survived all experimental tests
so far, we are sure that many details will need to be changed as we
gain more understanding.

The following is but some of the `items on the agenda' that come up
immediately to our mind: 
\begin{itemize}
\item to understand the dual transform;
\item to study further Higgs field as frame vectors;
\item to see if there is a geometric origin to the Yukawa terms;
\item to obtain a better picture of the middle-energy range;
\item to understanding further the  neutrinos.
\end{itemize}
No doubt more items will come up even before we start to tackle any of
the above.   We find this an exciting prospect.

\section*{Bibliography}
\hspace*{\parindent}The following book and articles list most of the
relevant references:
\begin{itemize}
\item Chan Hong-Mo and Tsou Sheung Tsun, {\it Some Elementary
   Gauge Theory Concepts} (World Scientific, 1993).
\item Chan Hong-Mo and Tsou Sheung Tsun, Nonabelian
   Generalization of Electric--Magnetic Duality---A Brief Review,
   invited review paper, hep-th/9904102,
   {\em International J.\ Mod.\ Phys.}\ {\bf A14}\,(1999)\,2139--2172.
\item Chan Hong-Mo and Tsou Sheung Tsun, The Dualized Standard Model
   and its Applicatins---an Interim Report,
   invited review paper, hep-ph/9904406,
   {\em International J.\ Mod.\ Phys.}\ {\bf A14}\,(1999)\,2173--2203.
\item Tsou Sheung Tsun, Concepts in Gauge Theory leading to 
   Electric--magnetic Duality
   (lecture course), in 
   {\em Proc.\ Summer School on Geometric Methods for Quantum Field 
   Theory}, Colombia, July 1999, 
   ed.\ H Ocampo, S Paycha and
   A Reyes, World Scientific, Singapore, 2000, hep-th/0006178.
\item Chan Hong-Mo, hep-th/0007016, Yang--Mills Duality and the 
   Generation Puzzle,
   invited
   lecture at the Intern.\ Conf.\ on Fund.\  Sciences, March 2000, 
   Singapore, {\em International J.\ Mod.\ Phys.}\ {\bf A16}\,(2001)\,163.
\end{itemize}

\end{document}